%% file: main.tex
\documentclass[%
 reprint,
 superscriptaddress,
 amsmath,amssymb,
 aps,
pre
]{revtex4-2}

\usepackage{graphicx}
\usepackage{bm}
\usepackage{hyperref}

\usepackage{multirow}

\begin{document}


\title{Employment and Land Use in the United States:\\ cross-sectional and short-term Trends reveal the Importance of the 96th Meridian}

\author{Diego Rybski}
\email{ca-dr@rybski.de}
\affiliation{Potsdam Institute for Climate Impact Research -- PIK, Member of Leibniz Association, P.O. Box 601203, 14412 Potsdam, Germany}
\affiliation{Department of Environmental Science Policy and Management, University of California Berkeley, 130 Mulford Hall \#3114, Berkeley, CA 94720, USA}
\affiliation{Complexity Science Hub Vienna, Josefstädterstrasse 39, A-1090 Vienna, Austria}

\author{David P. Helmers}
\affiliation{SILVIS Lab, Department of Forest and Wildlife Ecology, University of Wisconsin-Madison, Madison, Wisconsin, USA}

\author{Prajal Pradhan}
\affiliation{Potsdam Institute for Climate Impact Research -- PIK, Member of Leibniz Association, P.O. Box 601203, 14412 Potsdam, Germany}
\affiliation{Bauhaus Earth gGmbH, Berlin, Germany}

\author{Shade T. Shutters}
\affiliation{School of Complex Adaptive Systems, Arizona State University, P.O. Box 872701, Tempe, AZ 85287, USA}
\affiliation{Global Climate Forum, Neue Promenade 6, 10178 Berlin, Germany}

\author{Volker Radeloff}
\affiliation{SILVIS Lab, Department of Forest and Wildlife Ecology, University of Wisconsin-Madison, Madison, Wisconsin, USA}

\author{Van Butsic}
\affiliation{Department of Environmental Science Policy and Management, University of California Berkeley, 130 Mulford Hall \#3114, Berkeley, CA 94720, USA}

\date{\today}

\begin{abstract}
A pervasive trend in economic development is the shift from agricultural to manufacturing and finally to service economies. Because the type of employment associated with each sector influences natural resource use this global trend may manifest itself in predictable land use transitions. We relate these to changes in urban, agricultural, and natural lands. We find that the economic transition is common in most counties and across counties. Agricultural land expands to natural areas but is eventually replaced by urban. Relating sectors to land cover, we see a strong relationship between increases in service employment and the growth of urban areas. However, we find that both large agricultural areas and large natural areas also occur with high service employment. Finally, we test if the 100th Meridian is associated with predictable relationships between land cover and economic sectors and find strong results that suggest different development patterns in the East and West.
\end{abstract}

\keywords{Economic transitions, service sector employment, land cover change}
\maketitle


\section{Introduction}
The global economic shift from agricultural to industrial, and now service economies is one of the most prominent transformations of human history. As recently as 150 years ago, most workers in countries that are now considered economically advanced were farmers. Globally, today only about 28\,\% of all employment is in agriculture \cite{WorldBank2021} and in advanced economies this share is even less. For example, today less than 2\,\% of the United States population is employed in agriculture, although agricultural production continues to increase \cite{USDA2022}. This trend is not specific to the United States; most industrialized countries have undergone a similar economic transition \cite{LutzSRKR2013}. For most countries that are still industrializing, the transition away from agriculture employment and towards service and manufacturing will likely continue. However, although trends in employment are well known and documented, the connection of these trends to land use and land cover is less well understood.

The question of how land use is affected by the economic transitions is an important research gap because economic theory suggests that sectoral changes in employment should lead to predictable spatial patterns 
\cite{KrugmanP1991,Krugman1999}. Industries that require high agglomeration such as manufacturing and perhaps technology \cite{MorettiE2012}, should result in densely populated areas to take advantage of economies of scale \cite{SveikauskasL1975}. This suggests that as economies move to employ more people in manufacturing, these employees should live closer together, resulting in an increased urban area. As manufacturing jobs shift to service jobs, cities are likely to persist. Empirically, this effect can at least partially be observed in the rapid urbanization in most countries \cite{SetoFGR2011,UN2018}. Co-occurring with predictions of increased non-agricultural employment globally is the prediction of continued urbanization \cite{JiangN2017}. In the case of cities then, theories of economic geography nicely predict the global trends of urbanization we see today.

However, economic geography tells us less about what happens outside of cities to natural and agricultural lands. On the one hand, if reduced employment in agriculture is also related to reduced areas under production, it is possible that growth in cities might be associated with agricultural abandonment, and eventually the rewilding or regrowth of forests on formerly agricultural lands \cite{ChazdonLGCBC2020}. This may co-occur with the global trend of land protection \cite{ProtectedPlanet2020}. The phenomenon, known as the forest transition, has been observed in many countries where agricultural employment is decreasing, including the United States, Vietnam, and France \cite{MatherFN1999,MeyfroidtRL2010,LambinM2011}. On the other hand, urban areas still must receive food from agriculture. Depending on the population growth rate in the cities, their ability to import food, and the agricultural sector's ability to mechanize, agricultural land use may increase or decrease as city populations grow \cite{PradhanKCRBFK2020}. For example, even as cities have boomed in both population size and area in South America, the expansion of agricultural land use has also been pervasive there \cite{PhalanBBDSSB2013,PiquerRodriguezBBGGVMK2018}. Therefore, how natural and agricultural land uses change while economies shift towards service sector employment and city dwelling is unclear.

While economic theory has focused on the role of agglomeration in creating cities, geographers have long focused on the role of climate, soils, history, culture and governance on determining land use \cite{MeyfrodtP2016}. In the real world, it is likely that these forces help to shape the role of land use as countries and regions go through the economic transition. For example, many European countries have implemented policies to preserve farmland and establish strong boundaries between cities, villages, and agricultural areas. In the United States, such land use controls are less common, and suburban and ex-urban growth is widespread \cite{RadeloffHKMABMBHMSS2018}. Whether natural features also play a strong role in shaping land use outcomes during the economic transition is not well understood. While it is clear that the location of agriculture and natural areas are strongly influenced by the forces of geography, for example farmlands are typically on fertile soil, while natural areas are often those areas less habitable for humans \cite{JoppaP2009}, it is less clear how these land uses change in relationship to economic transitions.

In this study, we use data on employment and land cover in the United States to empirically explore the relationship between employment, sectoral change, and land cover. To start, we identify if there is a regular pattern of economic transition in the United States. We do this by evaluating county-level sectoral employment data in the United States from 2001 to 2016. We use this data to classify changes in sectoral employment. Next, we identify how land cover is related to total employment by combining employment and land cover data at the county scale and mapping the cross-sectional relationship between land cover and total employment. Third, we expand this analysis to break total employment into its sectors. We also analyze how individual sectoral changes are related to specific land cover changes at the county scale. Finally, using our results concerning land cover and economic transitions we explore the possibility that natural factors (climate in this case) still play a pervasive role in land cover, even as employment patterns converge.

\section{Methods}

\subsection{Study Area}
The study area for our analysis was the conterminous United States. We chose to study the United States because it has a high variation of employment and land use as well as natural and climatic features. Thus, it makes a robust study area to explore changes in employment and land use. There are also historical and contemporary arguments that distinct changes in rainfall around the 100th Meridian have greatly influenced land use and economic development \cite{PowellJW1890,SeagerLFTWNLH2018-1}. Within the United States, our unit of analysis was the county. Counties range in size from 59 sq km to 51,947 sq km. The number of employees within counties varies from 4.362 million (Los Angeles, CA) to less than 100 (Loving, TX) total employees. There are over 3000 counties in the conterminous United States, providing ample data points for our analysis.

\subsection{Data}
Our analysis was based on two primary data sources. First, we used data from the Bureau of Economic Analysis which estimates the number of employees in 32 employment categories for each year from 1969 to 2018 on the county scale \cite{BEA2021}. We then aggregated this data into three sectors (agriculture, manufacturing, and service employment) and calculated the shares of employment in each category by dividing each sector’s employment by the sum of employment. We excluded government employment from our dataset because it was unclear what sector these employees best fit. Therefore, our analysis is most precisely understood as an analysis of private sector employment. Out of the 3,109 counties in the conterminous United States, sufficient data exists in this dataset to estimate slope coefficients for sectoral employment for 3,007 counties. Missing data precludes using the remaining 102 counties.

Second, we used the National Land Cover Dataset –- NLCD \cite{MRLCNLCD2021}, land cover maps produced in 1991, 2001, 2004, 2006, 2008, 2011, 2013, and 2016. Because the 1991 NLCD is not directly comparable to other years, we exclude this period from our analysis. We, therefore, use NLCD data from 2001 to 2016 (7 samples). The NLCD was developed by classifying satellite imagery from Landsat satellites resulting in a land cover map with 21 land cover categories. We reclassified NLCD data into three land cover categories: natural lands (classes 41, 42, 43, 52, 71), agriculture (classes 81, 82), and urban areas (classes 21, 22, 23, 24). Other classes are not considered.

\subsection{Did shifts in employment and land cover follow similar patterns over time?}
We characterized the development of the three sectors (agriculture, manufacturing, and service shares) over time and, to be consistent, focused on the same period for land cover and employment (2001-2016). We applied linear regressions (ordinary least squares from the polyfit-function of the numpy-package in python) to obtain the slope of each sector independently. The dependent variable of these regressions was the share of employment in each sector, and time was the independent variable. We conducted the analysis at the county scale, such that the slope of each sector share was individually estimated for each county over our study period. The choice of using a linear regression instead of a more flexible functional form is justified by the relatively short time window in which more complex dynamics \cite{LutzSRKR2013,RybskiPSBK2021} can be approximated by linear behavior. Due to normalization, the slopes of the three sectors add up to zero, with only minor statistical deviations.

We followed a similar approach to estimate the slopes of land cover shares. First, to calculate the share of each land category in each county, we divided the amount of land in each of the reclassified categories by the land area of each county. We did this for each NLCD period used in our study. Next, using this share as the dependent variable, we used linear regressions to estimate the slope of land cover share changes from 2001 to 2016 at the county scale. As with employment, we used a linear model, which is especially appropriate given the limited number of times steps for land cover (n=7).

\subsection{How is land cover related to total employment?}
To understand how land cover is related to total employment and to check for power-law relationships, we combined employment data with land cover data and conducted a series of cross-sectional analyses. That is, we looked at these relationships across space instead of time. For this analysis, we focused on using data from 2016, our most recent year for land cover. For our first analysis, we binned total employment according to the natural logarithm, leading to 10 bins representing approximately 102 through 106 employees. We then calculated the mean share of land in agriculture, natural, and urban classes, across those counties in the respective bin. We plotted these mean values to show the average share of land in agriculture, natural, and urban for each bin of total employees.

\subsection{How are shifts in sectoral employment related to land cover?}
We used cross-sectional data (land cover and employment in 2016) to associate shares of land cover with shares of employment. To visually identify relationships, we created a 3x3 plot that associated the shares of each class of land cover with each class of employment. Finally, because service sector employment is the largest share in nearly all counties, and because the trend in nearly all counties is increasing levels of service employment, we built a table displaying the relationships between land cover shares and different levels of service employment.

\subsection{Are there patterns of land cover and employment shifts that suggest the influence of geography?}
There is a long history in the United States of defining East and West at the 100th Meridian, which generally follows the edge of a rainfall gradient [for a recent overview, we refer to \cite{SeagerLFTWNLH2018-1,SeagerFLTWNLH2018-2}. Generally, areas east of the 100th Meridian are wetter and more densely populated than the arid areas west of the 100th Meridian. Here, we tested to see if this gradient still holds for changes in contemporary land cover and for sectoral employment. The first step of this analysis was to identify the Meridian where discontinuities in land cover or employment were evident. We did this by running a series of regression discontinuity equations searching for the largest breakpoints between 104 degrees and 90 degrees in 1 degree steps. The dependent variable in these regressions was the slopes of land cover change and employment estimated previously. In total then, we ran 15 regressions for each of the six estimated slopes (agriculture, natural, and urban land cover and agricultural, manufacturing, and service employment) (90 regressions in total). For each regression, we used a bandwidth of 10 degrees. We then determined the best breakpoint by observing the coefficient estimates and standard errors for each regression and using an ad-hoc approach where the breakpoints with the biggest discontinuities were selected for further analysis.

Once the location that provided the strongest breakpoint was determined, we estimated the regressions for three different bandwidths -- 5 degrees, 10 degrees, and 30 degrees. We chose to use three bandwidths as there is limited theory to guide bandwidth selection and multiple bandwidths serve as a robustness check. By comparing the model output across land covers and sectoral employment, we were able to isolate relative changes around the Meridian. All regression discontinuity regressions were conducted using the rdrobust command in Stata \cite{StataCorp2021}. This command estimates a local polynomial regression with robust bias-corrected confidence intervals and p-values.

\section{Results}

\subsection{Did shifts in employment and land cover follow predictable patterns over time?}
\begin{figure*}
\centering
\includegraphics[width=1\textwidth]{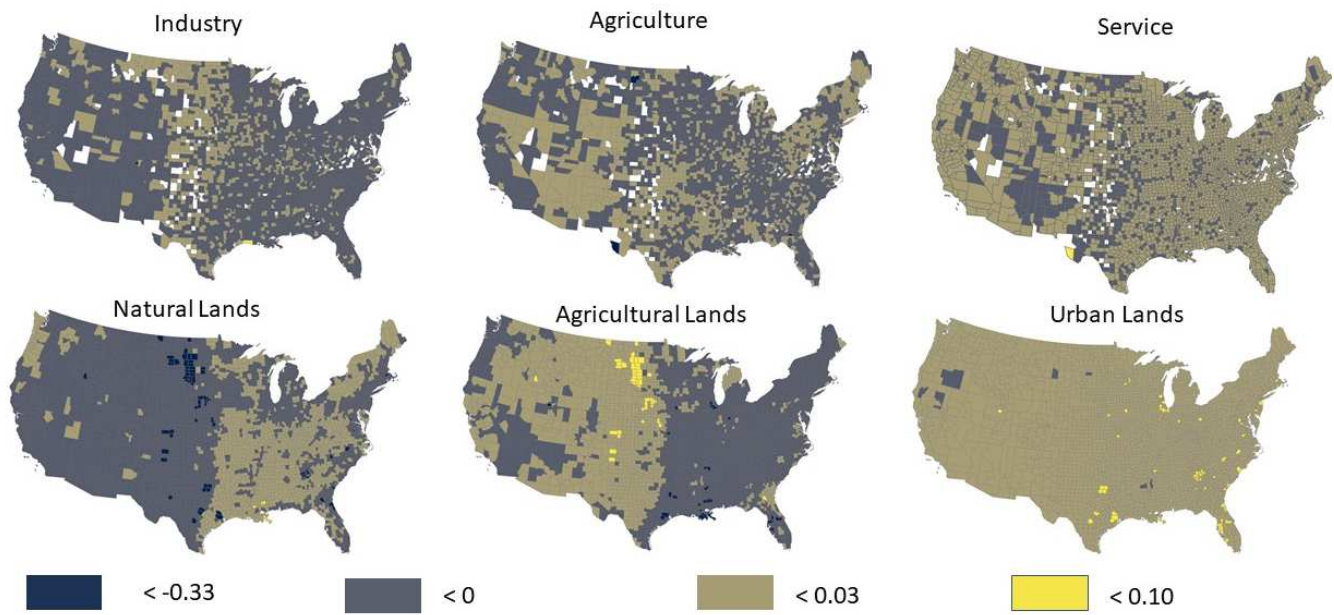}
\caption{Changes in economic sectors and changes in land cover in the United States. The maps depict changes in counties during the period from 2001 to 2016. The top row shows employment changes (agriculture, manufacturing, and service), and the bottom shows changes in land cover (natural, agriculture, and urban).}
\label{fig:fig1}
\end{figure*}

We found that shifts in sectoral employment followed the expected pattern over time. The share of service sector employment increased in 2413 out of 3116 counties (Fig.~\ref{fig:fig1}). In these counties, both agricultural and manufacturing employment fell in 1414 counties, only agricultural fell in 285, and only manufacturing fell in 714. Spatially, there were no obvious patterns in employment changes.

The land cover also changed in predictable patterns over time. In nearly every county, 3096 out of 3109, urban land areas increased from 1991 to 2016. Changes in agricultural and natural areas were more varied. Nearly 31\,\% of counties had increased agricultural areas, while 42\,\% had increased natural areas. These changes had a pronounced geographic pattern: while urban growth occurred all over the United States, growth in agricultural lands was pronounced in the West. In contrast, the growth of natural areas was pronounced in the East. Notable exceptions to these patterns occurred in Florida, Michigan, Nevada, and the Pacific Northwest. 

\subsection{How is land cover related to total employment?}
\begin{figure*}
\centering
\includegraphics[width=0.75\textwidth]{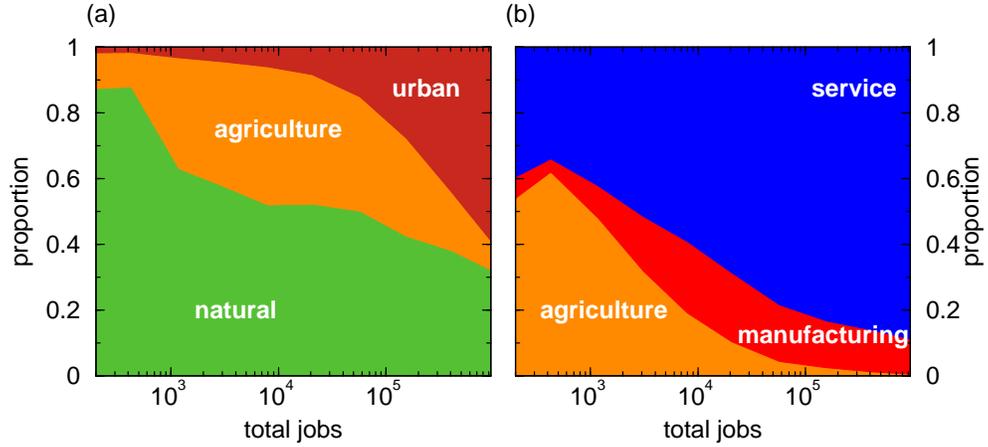}
\caption{Composition of land cover and employment as a function of total employment on the county scale in 2016. (a) The proportions of natural, agricultural, and urban land cover are represented in a similar way as in Foley et al. 2005. (b) The analogous representation for employment categories are agriculture, manufacturing, and service. The total employment is on a logarithmic scale, and bins according to the natural logarithm have been used to calculate the proportions. Only very few counties exhibit very small numbers of total employment -– accordingly, the decreasing agriculture at the left can be disregarded due to poor statistics.}
\label{fig:fig2}
\end{figure*}

\begin{figure*}
\centering
\includegraphics[width=0.75\textwidth]{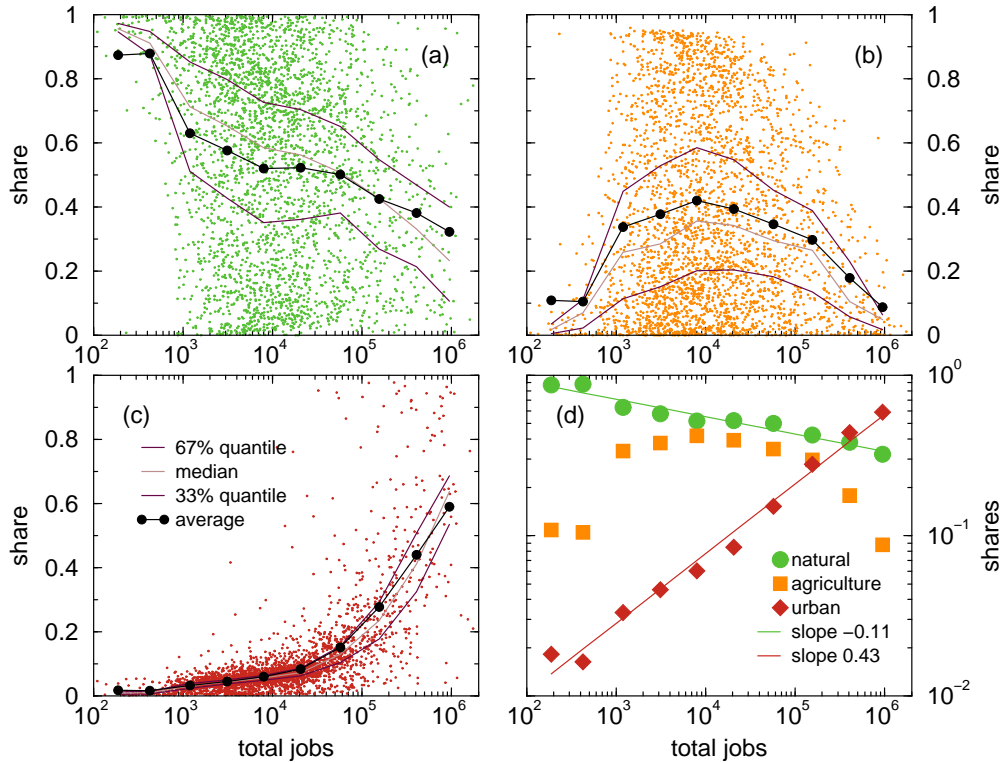}
\caption{Association between land cover shares and total employment in US counties (2016). The share of (a) natural, (b) agriculture, and (c) urban land is plotted vs total employment (log-scale). Dots represent individual counties. Based on logarithmic bins, filled circles represent averages, and brown lines represent upper and lower 1/3 quantiles. The light brown curve is the median. In contrast to Fig.~\ref{fig:fig2}, here, the shares have not been “stacked”. In (d), the average values are shown in a double-logarithmic representation in which a straight line corresponds to a power law. The solid lines represent linear regressions to the logarithmic values resulting in the slopes of -0.11 for natural and 0.43 for urban land cover. This means, on average, a county with ten times more employment exhibits (100.43)=2.7 times more urban land cover. Similarly, the share of natural land is (10-0.11)=0.8 times smaller -– but with considerable uncertainty as visible in (a).}
\label{fig:fig3}
\end{figure*}

\begin{figure*}
\centering
\includegraphics[width=0.65\textwidth]{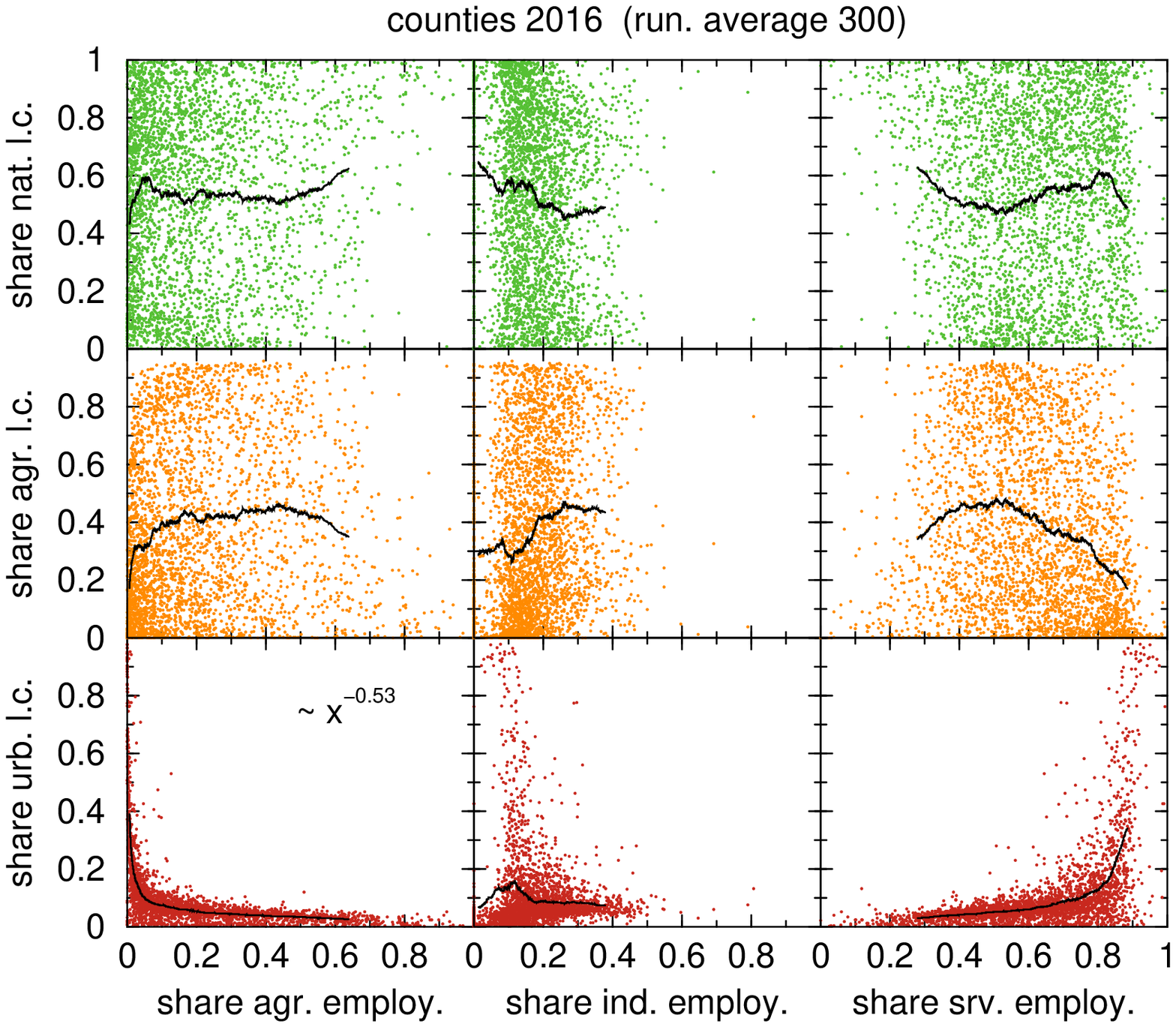}
\caption{Correlations between land cover shares and employment shares. In the nine scatter plots, each land cover class (natural, agriculture, urban) is compared to each employment class (agriculture, manufacturing, service). For example, in the top left, the share of natural land cover is related to the share of agriculture employment. Each dot represents a county in 2016, and the black curve stems from a running average (300 values). In the lower left, shares of urban land cover and agriculture approximately follow a power law with an exponent of -0.53.}
\label{fig:fig4}
\end{figure*}

In our cross-sectional analysis, as the number of employees increases, agricultural land first increases, generally at the expense of natural areas (Fig.~\ref{fig:fig2}). However, at higher levels of total employment, agricultural land is often overtaken by urban land. Natural land is replaced by agriculture at low levels of employment and, to a lesser extent, by urban development as the total number of employees increases (Figs.~\ref{fig:fig3} \& \ref{fig:fig4}). Urban lands are strongly and positively correlated with the total number of employees.

Our inspection suggests that natural lands approximately decrease with a power law relationship as the total number of employees increases [Fig.~\ref{fig:fig3}(d)]. To quantify such a power law relationship between the share of land cover and total employment, we ran linear regression on the relationship between the natural log of both total employment and land cover share. We then observed whether the regression slopes indicated a power law relationship. Likewise, the urban area increases with an approximate power law relationship as the total number of employees increases. As the agriculture curve is hump-shaped, no power law relationship can be expected.

\subsection{How are shifts in sectoral employment related to land cover?}
Relating employment in the three main sectors to land cover, we see a strong relationship between increases in service employment and the growth of urban areas. However, we find that both large agricultural areas and large natural areas are also possible with high service sector employment (Tab.~\ref{tab:tab1}). Additionally, we find that on average large shares of natural areas are more likely to be associated with high service sector employment than large shares of agricultural land cover. Indeed, when service sector employment hits 70\,\% (about the average in our dataset), only 30\,\% of counties have at least 50\,\% agricultural land. For the same level of service employment, over 60\,\% of counties have at least 50\,\% natural lands. 

\begin{table*}[htbp]
  \centering
  \caption{Shares of land cover in relation to service sector employment. Number of counties, out of total counties in parenthesis. Numbers in brackets are absolute counts out of total 3056 counties.}
    \begin{tabular}{|r|r|r|r|r|r|r|r|}
    \hline
    \multicolumn{2}{|r|}{\multirow{3}[4]{*}{}} & \multicolumn{6}{|c|}{Share of service} \\
    \cline{3-8}
    \multicolumn{2}{|r|}{} & 0.35 -- 0.44 & 0.45 -- 0.54 & 0.55-0.64 & 0.65--0.74 & 0.75 -- 0.84 & 0.85--0.95 \\
    \multicolumn{2}{|r|}{} & (346)   & (501)   & (558)   & (581)   & (538)   & (283) \\
     \hline
    Ag land cover greater & 0.25 (1661) & 0.67   & 0.68   & 0.64   & 0.53   & 0.44   & 0.27 \\
    than   & 0.5 (996) & 0.45   & 0.46   & 0.42   & 0.3    & 0.2    & 0.12 \\
           & 0.75 (470) & 0.22   & 0.27   & 0.21   & 0.14   & 0.07   & 0.02 \\
     \hline
    Natural land cover greater  & .25 (2315) & 0.73   & 0.69   & 0.73   & 0.79   & 0.81   & 0.72 \\
    than   & 0.5 (1743) & 0.5    & 0.51   & 0.53   & 0.61   & 0.64   & 0.48 \\
           & 0.75 (950) & 0.29   & 0.27   & 0.27   & 0.34   & 0.35   & 0.23 \\
    \hline
    \end{tabular}
  \label{tab:tab1}
\end{table*}

\subsection{Are there patterns of land cover and employment shifts that suggest the influence of geography?}
Testing for discontinuities between 104- and 90-degrees longitude, we found that the largest effect at the 96th Meridian for both the slope of agriculture and natural areas. As expected, natural lands increased in the East while agricultural lands increased in the West. We did not find a statistically significant discontinuity for urban lands. For employment, we found no significant discontinuity for 5 and 10 degrees on either side of the 96th Meridian. We did find statistically significant differences when a bandwidth of 30 degrees was used with larger positive changes in manufacturing employment to the West and larger positive changes in service employment to the East (Tab.~\ref{tab:tab2}). 

\begin{table*}[htbp]
  \centering
  \caption{Discontinuity at the 96th Meridian for slope of natural lands, agriculture, and urban lands and agricultural, manufacturing, and service employment.}
    \begin{tabular}{|r|c|c|c|c|c|c|}
        
    \multicolumn{7}{c}{Land  Cover}\\

    \hline
    \multicolumn{1}{|c|}{} & \multicolumn{2}{|c|}{5 degrees} & \multicolumn{2}{|c|}{10 degrees} & \multicolumn{2}{|c|}{30 degrees}\\
    \multicolumn{1}{|c|}{} & \multicolumn{2}{|c|}{(n east=487, n west=421)} & \multicolumn{2}{|c|}{(n east=979, n west=589)} & \multicolumn{2}{|c|}{30(n east = 2064, n west = 943)}\\
    \cline{2-7}
    \multicolumn{1}{|c|}{} & Coef.  & Robust p-value & Coef.  & Robust p-value & Coef.  & Robust p-value \\
    \hline
    Natural Lands & -0.0008** & 0      & -0.0009** & 0      & -0.0015** & 0 \\
    \hline
    Ag. Lands & 0.00079** & 0.005 & 0.0009** & 0 & 0.0015** & 0 \\
    \hline
    Urban Lands & 9.50E-05 & 0.696  & 5.30E-05 & 0.463  & 2.80E-05 & 0.617 \\
    \hline
    \multicolumn{7}{c}{Employment} \\
    \hline
    \multicolumn{1}{|c|}{} & \multicolumn{2}{|c|}{5 degrees} & \multicolumn{2}{|c|}{10 degrees} & \multicolumn{2}{|c|}{30 degrees}\\
    \multicolumn{1}{|c|}{} & \multicolumn{2}{|c|}{(n east=487, n west=421)} & \multicolumn{2}{|c|}{(n east=979, n west=589)} & \multicolumn{2}{|c|}{30(n east = 2064, n west = 943)}\\
    \cline{2-7}
    \multicolumn{1}{|c|}{} & Coef.  & Robust p-value & Coef.  & Robust p-value & Coef.  & Robust p-value \\
    \hline
    Man. & 0.00046 & 0.586 & 0.00068 & 0.712 & 0.00178** & 0.015 \\
    \hline
    Ag. & -0.00016 & 0.972  & 0.00049 & 0.617  & 0.00033 & 0.848 \\
    \hline
    Service & -0.0003 & 0.699  & -0.0012 & 0.899  & -0.0021* & 0.072 \\
    \hline
    \multicolumn{7}{l}{** indicates statistical significance at 95\% level, * indicates statistical significance at the 90\% level.}
    \end{tabular}
    \label{tab:tab2}
\end{table*}

Using the 96th Meridian as a breakpoint, we see similar relationships between sectoral employment and land cover, although mean values tend to differ.

\section{Discussion}
Economic shifts from agriculture to manufacturing and eventually service sector employment have occurred in many countries over the last 100 years. Yet, how this change is connected to land use remains relatively less well documented. This is an important research gap because the management of land use during this transition has important implications for food production, the environment, and overall quality of life. Here, we evaluate this relationship using panel and cross-section data for the United States.

First, we found that most counties in the United States followed the expected pattern of sectoral employment. Most counties gained shares in service sector employment and lost some employment in either agriculture or manufacturing. This pattern did not differ on the two sides of the 96th Meridian. This result suggests that sectoral changes in employment are nearly universal and that we might expect these changes to take place in many places globally, independent of geography. Because this trend is common across large variations in climate and geography, our findings suggest that economic and institutional factors may play the crucial role in this transition process.

We found power law relationships between total employment and natural and urban land area. As the total number of employees increased, the urban area increased, and the natural area decreased. This latter relation resembles the relation between population size and the area of cities \cite[and references therein]{BattyF2011}. Interestingly, agricultural land showed more quadratic relationships to employment with maximum agricultural land share at low, but not extremely low, populations. While these relationships suggest general patterns, we also found high heterogeneity at the level of the individual counties. Examples of counties with high urban and high agricultural or high natural areas persist. Therefore, beyond the clear expansion of urban areas, increases in the total number of employees can lead to multiple land cover outcomes. Understanding how policy, geography, culture, and other factors determine whether urban lands are associated with losses in agricultural or natural lands is an area of needed future research.

Land cover also followed predictable patterns across the United States. Nearly every county gained in the urban area. Given the persistent gains in service jobs everywhere in the United States, this result should be expected, especially at the time of our study, which occurred before the massive switch to remote work. There were pronounced patterns, however, for agriculture and natural lands. Areas east of the 96th Meridian saw increases in natural land -- probably forest regrowth on previously abandoned agricultural land -- and the West saw increasing agricultural use. This pattern was striking and suggests that 200 years after European settlement, the climate is still having a large influence on land cover decisions in the United States.

While economic transition is predicted by theory to occur from agriculture to manufacturing and then service, land cover succession is less deterministic. Natural areas can be transformed by agriculture but also directly through urbanization. Farmland can be abandoned, and transitioned to natural lands or be converted to urban areas. Broadly, urban areas tend to persist. Combining economic and land cover transitions, empirically, we mostly find correlations between service employment and urban land cover. The relationships between other stages in the economic transition and land cover transitions are less well-determined. One reason is that a positive relationship between natural lands and agriculture employment is unlikely if employment in agriculture is related to the area of agricultural land. Likewise, there is no distinct economic sector that is related to natural land cover (except minors like forestry or tourism). Accordingly, despite the sequential dynamics of both employment and land cover development, the relationships between the sectors are not a one-to-one match. This heterogeneity suggests strong roles for policy and geography in shaping how land cover changes with nearly inevitable changes in sectoral employment. 

As with any empirical analysis, our work is limited by data availability. For instance, our unit of analysis is the county, and the area of counties varies largely between East and West. We do not think this influences the conclusions, but smaller and more balanced spatial units could permit more robust analyses. However, employment information is not available in such units, which leaves room for follow-up work. Likewise, one of the fastest growing land use types in the United States is the wildland-urban interface (WUI). In our study, WUI areas are either considered urban or natural lands. Although we suspect that service sector jobs may be particularly well suited for WUI settings, we are unable to measure this in this manuscript. Adding a fourth land cover type would be an interesting natural extension to this work. 

Since starting this research in 2019, we have seen a massive shift in how employees work. During the covid 19 pandemic, work-from-home has become the norm for millions of service sector employees. With these employees free to live anywhere, it is unclear if they will choose urban settings or more rural locations. In this way, work-from-home may alter the relationship between the economic transition and land use. While the world has changed work practices, climate change is also continuing to change land use. Much of the arid west is suffering historic drought conditions. If these persist, it is natural to imagine the trend of expanding agricultural land use will end. Indeed, in California new water laws have reduced access to over-draft groundwater and will almost certainly lead to large farmland abandonment \cite{BourqueSAMPBAL2019}. Combined changes in working practices and climate may rearrange the patterns found here in the future.

\begin{acknowledgments}
D.\ Rybski thanks the Alexander von Humboldt Foundation for financial support under the Feodor Lynen Fellowship.
D.\ Rybski is grateful to the Leibniz Association (project IMPETUS) for financially supporting this research.
\end{acknowledgments}

%
%

\input{main.bbl}

\end{document}

%% file: main.bbl
%